\documentclass[
    reprint,
    preprintnumbers,
    superscriptaddress,
    nofootinbib,
     amsmath,amssymb,
     aps,
     prd,
    floatfix, longbibliography
    ]{revtex4-1}
    
    \usepackage{aas_macros}
    \usepackage{graphicx}
    \usepackage[caption=false]{subfig}
    \usepackage[usenames,dvipsnames]{xcolor} 
    \usepackage{pgfplots}
    \usepackage[utf8]{inputenc}
    \usepackage{color}
    \usepackage{hyperref}
    \usepackage[normalem]{ulem} 
    \usepackage{enumitem}
    \usepackage{comment}
    \usepackage{bm}
\hypersetup{
  breaklinks = true,
  colorlinks   = true, 
  urlcolor     = Blue, 
  linkcolor    = Blue, 
  citecolor   = Blue 
}
    \allowdisplaybreaks[1]
    \graphicspath{{figures/}}

\usepackage{float}
\usepackage{siunitx}
\newcommand{\Order}[1]{ {\cal O}(#1)}
\newcommand{\arccosh}{ {\rm arccosh}}
\newcommand{\kappath}{ k}
\newcommand{\rhob}{ \bar{\rho} }
\newcommand{\Pb}{ \bar{P} }
\newcommand{\cad}{ c_{\rm ad} }
\newcommand{\rhoK}{ \rho_{{\rm K} }}
\newcommand{\rhobK}{ \bar{\rho}_{{\rm K} }}

\newcommand{\uK}[2]{ U^{#1}_{\phantom{#1}#2}}
\newcommand{\PK}{ P_{{\rm K} }}

\newcommand{\PbK}{ \Pb_{{\rm K} }}
\newcommand{\phib}{\bar{\phi}}
\newcommand{\QK}{X}

\newcommand{\oxford}{Astrophysics, University of Oxford, DWB, Keble Road, Oxford OX1 3RH, United Kingdom}
\newcommand{\CEICO}{CEICO - FZU, Institute of Physics of the Czech Academy of Sciences, Prague}

\newcommand{\icg}{Institute of Cosmology \& Gravitation, University of Portsmouth, Portsmouth, PO1 3FX, United Kingdom}

\newcommand{\Mpc}{{\rm Mpc}}

\begin{document}

\title{The Cosmological Dipole in Tilted Anisotropic Universes}

\author{Alicia Martín}
\email{alicia.martin@physics.ox.ac.uk}
\affiliation{\oxford}
\author{Constantinos Skordis}
\affiliation{\CEICO}
\affiliation{\oxford}
\author{Deaglan J. Bartlett}
\affiliation{\oxford}
\author{Harry Desmond}
\affiliation{\icg}
\author{Pedro G. Ferreira}
\affiliation{\oxford}
\author{Tariq Yasin}
\affiliation{\oxford}

\begin{abstract}
{There is tentative evidence for a mismatch between the rest frames of matter and the cosmic microwave background, the ``quasar dipole anomaly''.}
We consider such a dipole in tilted anisotropic models, for a range of scenarios and sources: spatial curvature, cosmic heat flux, large scale electromagnetic fields 
and a Khronon field. Crucially, we determine the ancillary effects on other cosmological observables in each of these models and we show that, apart from the case of the Khronon field, it is unlikely that one can obtain a dipole with the amplitude that is being observed unless one considers additional exotica.
\end{abstract}
\maketitle

\section{Introduction}

Our current  model of the Universe assumes that space-time is, on the largest scales, homogeneous and isotropic.  It relies on a few key, extra assumptions: a period of super-luminal expansion in the early Universe (dubbed Inflation) and the existence of dark matter and dark energy (in the form of a cosmological constant, $\Lambda$). This model, the $\Lambda$ Cold Dark Matter model ($\Lambda$CDM) is extremely successful, accurately predicting observable phenomena on a wide range of scales \cite{dodelson2020modern}.

Intriguing anomalies, or tensions, have emerged as cosmological data has improved. 
One such tension is the existence of an anomalous 
{dipole in the number counts of distant sources across the sky, as indicated by both quasar catalogues and radio galaxy data. This is the ``Ellis--Baldwin test'' \cite{Ellis_Baldwin}, which, if the anomalous dipole is genuine, therefore appears to fail. The most precise of such inferences are \cite{Secrest:2020has,Secrest:2022uvx}, although the test has a history dating back to the early 2000s \cite{Blake_Wall}; for a review of the test and its current observational status see \cite{Secrest_colloquium}. This may derive from the special-relativistic transformation of source counts due to our motion with respect to them (brightening, increase in frequency and aberration), or alternatively from intrinsic clustering. Assuming a $\Lambda$CDM-like power-spectrum the latter is impossible \cite{Secrest:2020has} because the sources are at too high redshift for clustering to contribute significantly, suggesting the kinematic effect that the rest frame of these objects differs from that of the Cosmic Microwave Background (CMB).}

{In this interpretation, our velocity with respect to these objects is at least around twice as large as our velocity with respect to the CMB (although roughly aligned in direction), and the heliocentric frame velocities of the matter and CMB differ from each other at $\sim$5$\sigma$ significance \cite{Secrest:2022uvx, Dam, Bohme}.} 
According to these analyses, the statistical evidence for this anomaly is therefore unambiguous and calls into question the assumption of 
isotropy underpinning the cosmological principle and hence $\Lambda$CDM.
The Ellis--Baldwin test has developed into a cottage industry in the past couple of years, with a flurry of activity concerning its robustness to systematic effects, statistical methodology and datasets used (e.g. \cite{Abghari:2024eja, 2025arXiv251023769V, 2025A&A...697A.112W, 2025PhRvD.111l3547V, 2024MNRAS.535L..49V, 2025MNRAS.537....1O, 2025OJAp....8E.143M, 2025MNRAS.543.3229L, 2025arXiv251100822B}). While the jury is still out, there has been no convincing demonstration of a flaw in the claimed $5\sigma$ detection.

{Only a limited number of works have attempted to build cosmological models that realise such dipole in number count phenomenology.} These typically work either by relaxing the assumption of isotropy in the cosmological model \cite{Krishnan:2022qbv,Constantin:2022dtj}, introducing super-horizon matter perturbations to alter the distribution of distant sources across the sky \cite{2025PhRvD.112h3524K, 2022JCAP...10..019D}, or introducing an intrinsic dipole into the CMB so that the measured and kinematic dipoles do not coincide \cite{1991PhRvD..44.3737T, 1996PhRvD..53.2908L, 2022JCAP...10..019D}.
Once one considers more general models, one expects that there will be other observational consequences which might be constrained with current data.
So far there has been little attempt at studying the interplay between an anomalous dipole and other observables and how these might be used to constrain the former within the context of a specific mathematical model. It is the purpose of this paper to do so for a particular suite of models. {Although there are hints of intrinsic number count dipole contributions in addition to a global kinematic effect \cite{Wagenveld}, we focus here on the kinematic interpretation of a difference in the rest frames of matter and the CMB. This is compatible with the current data. We thus restrict our attention to modelling the $\beta \equiv \Delta v/c \approx 10^{-3}$ difference between the velocities of the matter and CMB frames.}

In this paper, we consider a simplified version of the problem. We construct a universe in which the matter frame, in which the quasars live, is boosted relative to the {CMB} frame. We do not concern ourselves with the additional problem that the observation of the dipole throws up: that the observers (us) are boosted relative to the matter frame {\it and} the {CMB} frame {\it along {approximately} the same direction as the relative boost between the matter and the cosmic rest frame}. 
This fact brings additional complications and makes any attempt at building such a model even more difficult. Thus, for our simplified model, we will focus on tilted, anisotropic models.

 Tilted anisotropic models are an extension of the standard, globally anisotropic, or {\it Bianchi}, models \cite{Ellis:1968vb,King:1972td,ellis2012relativistic}. The matter component is given a boost, {relative to the Bianchi background frame (which we take to be the CMB frame)}, along the direction of the global anisotropy. It is, in some sense, the most economical way of {supporting} an anomalous dipole {within a homogeneous cosmology.} There are stringent conditions that the space-time, or the energy-momentum of the constituents of such a universe, must obey but, as a result, there are clear knock on effects between the existence, and amplitude, of an anomalous dipole and other degrees of freedom, such as, for example, shear, curvature or the presence of extra fields. Shear plays a prominent role in anisotropic models as it is always non-zero and can be constrained through a variety of methods, most notably with the CMB \cite{Barrow:1985tda,Bunn:1996ut}.

This paper looks at a handful of universes which support an anomalous dipole. Given that the claimed {anomalous} dipole is {observed to be} small, $\beta\sim {\rm few} \times 10^{-3}$, we consider weakly anisotropic models and, in all cases we find a direct relationship between the dipole and other observables. We find that, for these models, it is difficult to have such an anomalous dipole, even of such small amplitude.

The paper is structured as follows. In Section \ref{formalism} we present the mathematical formalism of tilted anisotropic models which underpins our results. In Section \ref{models} we derive expressions for the dipole in terms of other potential observables for four model Universes: an open Universe, one with large scale heat flow, 
one with large electromagnetic fields, and one with a Khronon field. In Section \ref{constraints} we show how other cosmological constraints impact the amplitude of the dipole. In Section \ref{discussion} we discuss our findings. 

Throughout the paper we adopt a $(-,+,+,+)$   metric signature and assume $c=1$. An overdot denotes differentiation with respect to cosmic time $t$, and a prime denotes differentiation with respect to the spatial coordinate $z$. 
We use the symmetrization and antisymmetrization convention of tensor indices such that 
 $2 X_{(\mu\nu)} = X_{\mu\nu} + X_{\nu\mu}$ and  $2 X_{[\mu\nu]} = X_{\mu\nu} - X_{\nu\mu}$ respectively.
We will use $\Omega_0$ for the fractional energy density in all matter today. 

\section{The formalism}
\label{formalism}
Our starting point is to consider globally homogeneous but anisotropic models. These can be described in terms of the Bianchi metrics. We will focus on the anisotropic, Bianchi V metric:
\begin{align}
    ds^2=-dt^2+a^2\left[e^{-2(b+ \alpha z)}(dx^2+dy^2)+e^{4b}dz^2\right], 
\label{eq:metric}
\end{align}
where $a$ and $b$ are functions of cosmic time and $\alpha$ is  a constant \cite{Ellis:1968vb,ellis2012relativistic}. The Bianchi V model is spatially hyperbolic (i.e. it is open) and in the 
limit $\alpha\rightarrow0$ we obtain a spatially flat, Bianchi I metric. We can define a Hubble expansion rate $H\equiv{\dot a}/a$ and a shear $\sigma\equiv3{\dot b}$.

The dynamics of such a universe are given by the Einstein field equations (note that $T_{11}=T_{22}$):
\begin{align}
3H^2 - \frac{\sigma^2}{3} - \frac{3\alpha^2e^{-4b}}{a^2} =& \kappa T_{00},
 \\
{2\alpha}\sigma =& -\kappa{ T_{03}},
 \\
\dot{\sigma} + 3H \sigma =&  \frac{\kappa}{a^2} \left( e^{-4b} T_{33}- e^{2\alpha z + 2b} T_{22}  \right) ,
 \\
\dot{H} + 3H^2 - 2\frac{\alpha^2 e^{-4b}}{a^2} =&  
\frac{\kappa}{2} T_{00} 
-  \frac{\kappa e^{2\alpha z+2b}}{3a^2}  T_{22}
\nonumber
\\
&
- \frac{\kappa  e^{-4b} }{6a^2}T_{33}, 
  \end{align}
where $\kappa \equiv 8 \pi G$  and $T_{\mu\nu}$ is the energy-momentum tensor.
For a perfect fluid the energy-momentum tensor takes the form
\begin{align}
T_{\mu\nu}^{(f)}=&(\rho+P) u_\mu u_\nu + P g_{\mu\nu},
\label{T_fluid}
\end{align}
where $\rho$ is the volumetric energy density and $P$ is the pressure of the fluid. The vector field $u^{\mu}$ is the fluid four velocity, that has components 
\begin{equation}
u^{\mu}= \left( \cosh\beta,0,0, \frac{\sinh\beta}{ae^{2b}} \right)
\approx \left(1,0,0, \frac{\beta}{ae^{2b}} \right) .
\label{eq:four_velocity}
\end{equation}
The boost parameter, $\beta\simeq v\ll 1$ is the dipole we are interested in.

Thus, $T_{\mu\nu}^{(f)}$ has components 
\begin{align}
T_{00}^{(f)} =& \rho, \\
T_{22}^{(f)} =& T_{11}^{(f)} =  P e^{-2\alpha z} a^2e^{-2b}, \\
T_{33}^{(f)} =&  P a^2e^{4b}, \\
T_{03}^{(f)} =& -(\rho + P) ae^{2b} \beta. 
\end{align}
plus corrections of $\Order{\beta^2}$.
From the covariant conservation of the  perfect fluid energy-momentum tensor $\nabla_{\mu} T^{(f)\mu \nu } = 0$ we obtain two independent equations:
\begin{align}
\dot{\rho} + (\rho + P) \left(3H   - 2\frac{\alpha e^{-2b}}{a}\beta\right)  =& 0 + \Order{\beta^2}, 
\\
  \dot{\beta} + (1 - 3 \cad^2)  H \beta 
+ \frac{2\alpha \cad^2 e^{-2b}}{a}\beta^2
+ \frac{2}{3}\sigma \beta 
=& 0+  \Order{\beta^3} , 
\end{align}
which govern the evolution of $\rho$ and $\beta$. The pressure is commonly given in terms of the equation of state parameter $w\equiv P/\rho$, 
and the adiabatic sound speed is defined via $\cad^2\equiv \dot{P}/ \dot{\rho} = w - \frac{\dot{w}}{3 H (1+w)}$. If $\dot{w}=0$ then $\cad^2=w$.

We will be considering a universe in which $\beta\simeq {\rm few} \times 10^{-3}$. This means that it will have a negligible impact on the background evolution and thus we are free to work with the standard Friedman equations for the Hubble rate and the background energy density. The dipole is, nevertheless, much larger than cosmological perturbations on large scales (which are of order $10^{-5}$) which means we won't unleash the full machinery of linear cosmological perturbation theory apart from when they include the anisotropic degrees of freedom or when they are useful to identify the nature of key, small quantities (as in the case of the Khronon field). Thus, in what follows, we will focus on the off-diagonal Einstein field equation (the ``$03$'' equation) and the traceless part of the spatial part of the equation (which gives us the evolution of the shear, $\sigma$).

\section{Sources of Tilt}
\label{models}
We are now interested in finding ways of generating  a (or constructing a self consistent) universe
with a dipole.  From the ``$03$" Einstein field equation, in the $\beta\ll1$ case
we have that

\begin{align}
{2\alpha}\sigma \simeq& -\kappa\left[{ T_{03}}  - ae^{2b}(\rho + P)\beta \right],
\label{dipolecondition}
\end{align}
where $T_{\mu\nu}$ is any of the additional sources of energy momentum to the boosted matter energy momentum tensor. From \eqref{dipolecondition} we see that there are two ways of maintaining a dipole. We either have $\alpha\neq 0$ and no additional energy momentum or, for $\alpha=0$, an additional source, $T_{03}\neq0$ to compensate for the dipole in the boosted matter component. 

It is intriguing to note that the presence of curvature allows for the presence of a preferred direction (the boost direction) in a way that the flat Universe does not \cite{ellis2012relativistic,Krishnan:2022qbv}. One possible way of seeing this is in the fact that the isometry algebra of Bianchi V is parametrized by an additional vector compared to that of Bianchi I and thus has less symmetry between the orthonormal basis' \footnote{To be specific, while the structure constants for the (Euclidean) Bianchi I model takes the form ${\cal C}^{c}_{\phantom{c}ab}=0$ while for the (open) Bianchi V we have that ${\cal C}^{c}_{\phantom{c}ab}=\delta^{c}_{\phantom{c}b}A_a-\delta^{c}_{\phantom{c}a}A_b$ where ${\vec A}$ is a constant vector.}. Another way of seeing this is that one can foliate the hyperbolic space in terms of flat 2-surfaces, $R^2$, leaving the additional dimension to soak up the curvature. 

While we could consider superpositions of these various effects -- curvature and the additional sources of anisotropy -- we will now look at each one in turn.

\subsection{Curvature}

Consider a Bianchi V model with $\alpha \neq 0$ and the energy tensor given by the perfect fluid energy tensor, $T_{\mu\nu} = T_{\mu\nu}^{(f)}$ \cite{Krishnan:2022qbv}. The relevant Einstein 
equations become
\begin{align}
2\alpha \sigma =& ae^{2b}\kappa (\rho + P) \beta, \\
\dot{\sigma} + 3 H\sigma =& \kappa ( \rho + P) \beta^2.   
\end{align}
plus higher $\beta$ corrections.
If $\sigma = 0$ the equations reduce to those of a usual open homogeneous and isotropic universe with untilted matter.

Since $\beta\ll1$ we can attempt to solve the equations where the anisotropic terms are small perturbations around a homogeneous and isotropic universe. 
The five unknowns are  $\rho$, $P$, $\beta$, $a$, $\sigma$ and we will assume that  $P = w \rho$ where $w$ is a time dependent equation of state. We can solve for $\rho$ and $P$ to get 
\begin{align}
\rho =& \frac{\rho_0}{a^3} e^{-3 \int{H(t)w(t)dt}}, 
\end{align}
and solving for  $\beta$ we have 
\begin{align}
\beta \simeq \frac{2\alpha}{(1+w)}\frac{\sigma}{ a\kappa \rho}. 
\end{align}

Note that the curvature parameter $\alpha$ can be re-expressed as
$\alpha = k^{1/2} =  H_0 \Omega^{1/2}_{k} $, where $\Omega_k>0$ is the usual fractional energy density in curvature, or curvature parameter, of a homogeneous and isotropic universe. 
We can also use the Friedman equations to get $3 H^2 \Omega = \kappa \rho$, and let $\Omega_0 = \Omega(z=0)$,
to find  an expression for the dipole today:
\begin{align}
\beta \simeq \frac{2}{3}\frac{1}{1+w}\frac{\Omega^{1/2}_{k}}{\Omega_0} \frac{\sigma_0}{H_0}.
\label{beta_today_curvature}
\end{align}
Thus, we have obtained an expression for the dipole in terms of other cosmological observables: the normalized shear today $\sigma_0/H_0$ and the curvature parameter $\Omega_k$.

It is instructive to see how the shear will evolve. Again, considering the case where $\beta\ll 1$, we have that
\begin{align}
{\dot \sigma}+3H\sigma-\frac{4\Omega_k H_0^2}{3a^2\Omega(1+w)H^2}\sigma^2  = 0,  
\end{align}
which has a formal solution:
\begin{align}
\frac{\sigma}{H}=\frac{1}{a^3}\left[C-\frac{4\Omega_kH^3_0}{3}\int\frac{d\ln a}{a^5(1+w)H^3 \Omega}\right]^{-1},
\end{align}
where $C$ is an integration constant. 
There are, therefore, corrections to the standard shear evolution ($\sigma \propto a^{-3}$) due to the presence of the dipole which may be relevant, depending on the amount of curvature and size of $\beta$.

\subsection{Heat flow}
An interesting possibility is to have heat flow due to one or more particle species or fields. 
Relativistic non-equilibrium thermodynamics has a long history, starting from the pioneering work of Eckart~\cite{Eckart:1940te} and of Landau and Lifshitz~\cite{LandauLifshitzVol6}.
These early approaches were plagued by instantaneous propagation of heat, and as was shown later, also of instability of fluctuations~\cite{Hiscock:1983zz}.
Based on relativistic kinetic theory, Israel and Stewart~\cite{Stewart1977,IsraelStewart1979,Israel:1979wp}  developed a phenomenological model
 describing heat and dissipation which brought into play several additional coefficients into the resulting equations.
  As shown in~\cite{Hiscock:1983zz}, the Israel-Stewart theory is causal, stable and has a well-posed initial value problem. Further formulations were
later proposed~\cite{AnderssonComer2010,Lopez-Monsalvo:2010oeo} based on a two-component Lagrangian which brings in entropy as a further dynamical component 
to the fluid number density. The reader is referred to~\cite{Andersson:2006nr} for a thorough review of all approaches.

Here we take as a starting point the Israel-Stewart theory and work within the so-called Eckart frame, where the energy-momentum tensor of the fluid takes the form
\begin{align}
T^{(f,Q)}_{\mu\nu}= (\rho+P) u_\mu u_\nu + P g_{\mu\nu} + 2 q_{(\mu} u_{\nu)}
\label{T_Q}
\end{align}
and differs from \eqref{T_fluid} through the last term $2 q_{(\mu} u_{\nu)}$, where the vector field $q_\mu$ obeys $u^\mu q_\mu=0$. 
The Eckart frame is also called the particle frame, as the fluid $4$-velocity $u^\mu$ is proportional to
the fluid particle flux. 
We work within the small dipole approximation and so frame transformations leave $\rho$ and $P$ invariant to lowest order in $\beta$, 
however,  $q_\mu$ is not; see \cite{Kopp:2016mhm} for an exposition.  
We ignore viscous stresses as we can consistently set them to zero within the Israel-Stewart theory, and moreover, they are invariant under a frame transformation to lowest order in $\beta$.
Note that the formulations of~\cite{AnderssonComer2010,Lopez-Monsalvo:2010oeo} also introduce a term proportional to $q_\mu q_\nu$ in \eqref{T_Q}, however, this term is
necessarily of higher order in $\beta$ and will not play a role in our discussion. To summarize, to lowest order in $\beta$ all aproaches converge to the case
we consider here.

In an FLRW Universe with a single fluid, one can always choose a frame where $q_\mu$ vanishes.
However, in the real Universe, we deal with several species, and thus we choose 
a non-zero $q_\mu$ for at least one of them. Within our metric ansatz \eqref{eq:metric} we align the spatial part of $q_\mu$ with the
 $z$-direction, such that $q_\mu = (q_0, 0, 0, Q)$.  Imposing $u^\mu q_\mu=0$ and working in the  small dipole approximation  on 
   a pure FLRW background so that $b=0$, this implies that $q_0 = - Q\beta / a$,  where $Q=Q(a)$.
Letting $T^{(f,Q)}_{03}=0$ to maintain an FLRW solution, we find that
\begin{align}
 Q = -  a (\rho + P)\beta,
\label{Q_beta}
\end{align}
and hence $Q$ is of order $\beta$. It turns out that the remaining components of the term $2 q_{(\mu} u_{\nu)}$ are of order 
$\beta^2$ or higher and so we can safely set them to zero. This means that standard energy-momentum conservation holds,
$\dot{\rho} + 3 H (\rho + P) = 0$.

The Israel-Stewart theory supplies us with a heat transport equation, which is a general relativistic version of the Maxwell-Cattaneo equation~\cite{Andersson:2006nr}.
From~\cite{Israel:1979wp} we have that
\begin{align}
 q_\mu = - k q_\mu^{\;\;\nu} \left( \nabla_\nu T + T a_\nu + \bar{\beta}_1 T  u^\rho\nabla_\rho q_\nu \right),
\label{Heat_transport}
\end{align}
where $q_{\mu\nu} \equiv u_\mu u_\nu + g_{\mu\nu}$ is the usual projector tensor, 
$a_\mu \equiv u^\nu \nabla_\nu u_\nu$ is the acceleration, $k$ is thermal conductivity and $T$ is the temperature.
The coefficient $\bar{\beta}_1$ is a phenomenological quantity and a function of $\rho$ and $P$ in general~\cite{Israel:1979wp}. 
Considering the $\mu=0$ and $\mu=3$ components of \eqref{Heat_transport} we find respectively
\begin{align}
 Q =&  -k  \left[ T' +  a T \left(\dot{\beta} + H   \beta \right)   + \bar{\beta}_1 T  \left(\dot{Q} - H Q  \right) \right],
\\
 Q =& -a  k \beta \dot{T} - k  \left[  T' + a T  \left(\dot{\beta} + H   \beta \right)  + \bar{\beta}_1 T  \left(\dot{Q} - H Q  \right)   \right],
\end{align}
where a prime denotes differentiation with respect to $z$.
 Now letting $\beta\rightarrow 0$ and given \eqref{Q_beta}, we would obtain
$T' = 0$, that is, if there was no dipole and heat flow, then there would not be any temperature difference. Thus, consistency requires that we assume that $T$ is of order $\beta$, and to lowest order we obtain the simple relation
\begin{align}
  a (\rho + P)\beta  \approx&   k  T', 
\label{Fourier_law}
\end{align}
which relates the dipole to the temperature gradient along its axis. Thus, to lowest $\beta$ order, the coefficient $\bar{\beta}_1$ plays no role. 
Note that \eqref{Fourier_law} is identical to what we would have obtained directly from
the Eckart theory~\cite{Eckart:1940te}; this is consistent within our small $\beta$ approximation.

Before closing this subsection, we consider the shear $\sigma$ which will be generated as a correction to FLRW dynamics produced by $\beta$.  We now find that
\begin{align}
\dot{\sigma} + 3 H \sigma =& -\kappa \rho (1 + w)\beta^2, 
\end{align}
{where the leading-order contribution is now $\mathcal{O}(\beta^2)$. Although we neglected $\mathcal{O}(\beta^2)$ corrections in previous equations (because those quantities already had nonzero $\mathcal{O}(\beta)$ terms) here the shear is sourced only at second order, so the $\mathcal{O}(\beta^2)$ term is the leading nonvanishing contribution and must be retained.}
This can be formally integrated to give
\begin{align}
\sigma=& \frac{1}{a^3} \left[ C - \kappa  \int dt  a^3 \rho (1+w) \beta^2 \right].  
\end{align}
Note that we now have a correction which is heavily suppressed relative to the homogenous term and which can decay but may grow depending on the behaviour of $\rho$ (through the choice of $w$).

\subsection{Electromagnetic Field}

We now extend the case of a universe containing a perfect fluid, such that it also contains an electromagnetic (EM) field, which will be responsible for the dipole. We set $\alpha=0$ (the Bianchi~I limit). The Maxwell tensor, $F_{\mu\nu}$ the EM field
will have an energy momentum tensor of the form 
\begin{align}
T^{\rm EM}_{\mu\nu}=\frac{1}{\mu_0}\left(F_{\alpha \mu}F^{\alpha}_{\phantom{\alpha}\nu}-\frac{1}{4}g_{\mu\nu}F_{\alpha\beta}F^{\alpha\beta} \right) ,
\label{eq:T_EM}
\end{align}
where $\mu_0$ is the permeability of the vacuum. The EM tensor $F_{\mu\nu}$ may be written in terms of  electric $E^i$ and magnetic $B^i$ fields 
by identifying $F_{0i} = g_{ij} E^j $ and $F_{ij} = \epsilon_{ijk} B^k$.
The vectorial nature of the electromagnetic field make it particularly suitable for supporting global anisotropy \cite{1980ApJ...235..307D,1980PhLA...79...19L,1981GReGr..13..795L}.

Computing the components $T^{(\rm EM)}_{0A}$ and $T^{(\rm EM)}_{3A}$ and ensuring that they vanish as required by our metric ansatz
is possible if, and only if,  $E_3 = B_3 = 0$. This allows us, without loss of generality, to choose the electric field to be solely in the $x$-direction: $E^i = (E,0,0)$. Computing now the $T_{03}^{(\rm EM)}$ that is required for having a dipole, we find that only the $B_2$ component of the magnetic field may contribute. This means $B_1$, if present, will play no role in having a dipole. Furthermore, from the Maxwell equation $\nabla_{[0} F_{23]}=0$ we find that $B_1$ decoupled from the other variables,
and hence, without loss of generality, we may set it to zero so that $B^i = (0, B , 0)$. 

Computing $T^{(\rm EM)}_{AB}$, which should be diagonal,  as required by our metric ansatz, gives the additional constraint $E = \pm  B$. Incidentally, this leads to
 $T^{(\rm EM)}_{11} =  T^{(\rm EM)}_{22} = 0$. The remaining Maxwell equations, $\nabla_{[\mu} F_{\nu\rho]}=0$ and 
 $\nabla_\mu F^{\mu\nu} = \mu_0 J^\nu$, for current $J^\mu$ imply that $J^\mu = 0$ (this setup is not supported by a current) and that
\begin{align}
 B = \frac{B_0}{a^3}, 
\end{align}
 where $B_0$ is a constant. 

 Having found the solution for the electric and magnetic field in our setup and assuming once more that $\beta $ is small (and that $\sigma \sim \Order{\beta} $ or higher) we find the dipole
 \begin{align}
 \beta = \frac{1}{ \mu_0 a^4(\rho+P)} B_0 {E_0}, 
 \end{align}
 Today, this gives the explicit expression
\begin{equation}
\beta=  \frac{1}{\mu_0 (1+w) \rho_0}  B_0 E_0,
\end{equation}
where $E_0 = B_0$, but we have kept the dependence on each field component separate so that we can capitalize on bounds on either, when looking at the ancillary constraints.
We stress that it is not possible to source a dipole without having both an electric and a magnetic field.

The remaining non-zero components of the EM energy momentum tensor are $a^2 e^{4b} T_{00} =  \frac{1}{\mu_0} a^4 e^{2b} B^2 = - T_{33}$ so the EM energy density
scales as $a^{-4}$, as expected. We note, in particular, that the EM field will contribute to the Friedmann equation. 
As in the previous examples, we finally look at the shear. We find
that, to leading order in $\beta$, we have
\begin{align}
\dot{\sigma} + 3 H\sigma =&  - \kappa \frac{1}{\mu_0} \frac{B_0^2}{a^4},   
\end{align}
which we can integrate to get
\begin{align}
\sigma =& \frac{1}{a^3} \left( C - \frac{\kappa B_0^2 }{\mu_0}  \int^t \frac{dt'}{a} \right).  
\end{align}
In this case, we have an additional term which decays more slowly than the homogeneous term, albeit suppressed by factors of $B_0^2$.

\subsection{Khronon field }

One possible source of anisotropy is a Khronon field, i.e. a scalar field which is used to define a preferred time-like spatial foliation of the space-time. 
Khronons have been invoked to construct non-relativistic theories of quantum gravity and, more recently, as ingredients in relativistic formulations of modified Newtonian 
dynamics~\cite{Blanchet:2024mvy} due to their ability to emulate cosmological dynamics of dust~\cite{Scherrer:2004au}.
They also form an integral part of ghost condensates~\cite{Arkani-Hamed:2003pdi} (which contain additional terms as dictated by effective field theory) 
which  have also been considered in the context of dark matter~\cite{Furukawa:2010gr}.

We will, now, consider a theory with a Khronon field as well as the normal matter fluid. The action for the Khronon scalar field $\phi$ has the form
\begin{align}
S = \frac{1}{8 \pi G}\int{ d^4x \sqrt{-g} F(\QK)}, 
\end{align}
where $\QK = - g^{\mu \nu} \nabla_{\mu} \phi \nabla_{\nu} \phi$. We set the function $F(\QK) = \mu^2(\QK-1)^2 + \ldots$, where the ellipsis denotes higher order terms in this Taylor expansion that play a role in the early Universe. In the late Universe, the lowest order term of this specific form
leads to dark matter behaviour for the Khronon. The Khronon field $\phi$ is assumed to have units of time
so that $\QK$ is dimensionless and $\mu$ has units of inverse length. Note that we have defined $\QK$ 
such that it is manifestly positive on a Friedman background.

The Khronon contributes an energy momentum tensor of the form
\begin{align}
 T^{(\rm K)}_{\mu\nu} =& \frac{1}{ \kappa } \left[  2F_{\QK}  \nabla_{\mu} \phi \nabla_{\nu} \phi  +  F g_{\mu\nu}  \right], 
\label{T_Khronon}
\end{align}
where $F_{\QK} \equiv \partial F / \partial {\QK}$.
The similarity of \eqref{T_Khronon} to that of the energy-momentum tensor of a perfect fluid is suggestive. Indeed, we may define a Khronon velocity by
\begin{align}
 \uK{}{\mu} \equiv  \frac{\nabla_\mu \phi}{\sqrt{\QK}}, 
\end{align}
so that $T^{(\rm K)}_{\mu\nu}  = \left(\rhoK + \PK\right)\uK{}{\mu} \uK{}{\nu}  + \PK g_{\mu\nu}$. This allows us to identify the energy density
$\kappa \rhoK \equiv 2 \QK F_{\QK} - F$,  pressure $\kappa \PK \equiv F$ and adiabatic speed of sound  $\cad^2 = F_{\QK} / ( F_{\QK} + 2 {\QK} F_{\QK\QK} )$, such that $dP/d\rho = \delta P /\delta \rho =  \cad^2$.

For our choice of function, the scalar field equation of motion is $\nabla_\mu[({\QK} - 1)\nabla^{\mu}\phi] =  0$, and on FLRW we find
\begin{align}
  \partial_t \left[ a^3 (\QK-1)   \dot{\phi}\right]  
= a  \partial_z \left[  (\QK-1) \phi'\right].
\label{EOM_Khronon}
\end{align}
The task is now to solve the above equation and in the FLRW case where $\phi=\phib(t)$ and $\QK= \dot{\phib}^2$ this
can be done exactly. However, in the late Universe it may be shown that $\dot{\phib} \approx 1 + I_0/(2 a^3) + \ldots$ for an initial condition $I_0$, while 
 the energy density and equation of state of the Khronon approximate to $\kappa \rhobK\approx  4 \mu^2 I_0 /a^3 + \ldots$ 
and $w_K = \PbK/\rhobK \approx  I_0/ (4 a^3)+ \ldots$ respectively, where a `bar' denotes FLRW background quantities. This is the well-known result that the Khronon
approximately contributes a dust (or dark matter) component in the Universe. It is thus sufficient to set $w_K = \cad^2 \approx 0$ in the late Universe.

To have a dipole, we need a $z$-dependence in the Khronon and we do this perturbatively as $\phi(t,z) = \phib(t) + \varphi(t,z)$,
 where $\varphi\ll \phib$.  In the limit $w_K = \cad^2 \approx 0$, the energy-momentum tensor of the Khronon including the perturbations is
\begin{align}
T^{(\rm K)}_{00}  = \rhoK  = \rhobK (1 + \delta_K), 
\\
T^{(\rm K)}_{03}  = \rhobK  \varphi', 
\\
T^{(\rm K)}_{33}  = {\varphi'}^2 \rhobK. 
\end{align}
while $T^{(\rm K)}_{11} = T^{(\rm K)}_{22}  \approx 0$.
Since for FLRW $T_{\mu\nu}$ cannot depend on $z$, this implies that neither $\delta_K$ nor $\varphi'$ can depend on $z$. 
We relate $\delta_K$ to $\varphi$ using the exact equation 
\begin{align}
\QK =  \dot{\phi}^2  - \frac{1}{a^2}   (\varphi')^2,
\label{Q_exact}
\end{align} 
and $\delta_K = \delta \rho_K / \rhob_K = ( F_{\QK} + 2 \QK F_{\QK\QK}  ) \delta \QK / \rhob_K$  to get
\begin{align}
\delta_K =   \frac{2(3 \QK-1)}{   (\QK-1)(3\QK+1)  }  \left[ 2 \dot{\phib} \dot{\varphi} + \dot{\varphi}^2  - \frac{1}{a^2}   (\varphi')^2 \right].  
\label{delta_K}
\end{align}
Notice that it would seem that $\delta_K$ blows up as $\QK\rightarrow1$, however, this only means that $\dot{\varphi}$ and $\varphi'$ 
should vanish in that limit to keep $\delta_K$ finite.
Importantly, requiring that $\delta_K$  is a function of $t$ only, and that $\varphi'$ is independent of $z$, then  \eqref{delta_K} results in
$\dot{\varphi}$ also being independent of $z$. Hence, the only possibility is that $\varphi = \varphi_0(t) + \varphi_1 z$ where $\varphi_1$ is a constant
and we may further absorb $\varphi_0(t)$ into the background $\phib(t)$ without loss of generality. To conclude, we seek solutions of the form
 $\phi = \phib +  \varphi_1 z$ to generate a dipole.

With the above ansatz for $\phi$ the RHS of \eqref{EOM_Khronon} vanishes and we get
\begin{align}
  \partial_0 \left[ a^3 (\QK-1)  \dot{\phi}\right]  = 0. 
\label{Khronon_EOM}
\end{align}
Integrating  and using \eqref{Q_exact} leads to the cubic equation
\begin{align}
  \left( \dot{\phi}^2  - 1 - v^2 \right)  \dot{\phi} = \frac{I_0}{a^3}, 
\end{align}
where $I_0$ is an initial condition and $v\equiv \varphi_1/a$.
We can solve the above equation exactly as
\begin{align}
\dot{\phi} = 2 \sqrt{\frac{1+v^2}{3}}\cosh\left[ \frac{1}{3}\arccosh\left( \frac{3\sqrt{3}I_0}{2a^3(1+v^2)^{3/2}}\right)\right], 
\end{align}
where we have taken the positive solution. If $v=0$, we recover the exact FLRW solution for $\phib(t)$, and morover, expanding around $\dot{\phib} = 1$ recovers 
the late Universe solution discussed above. For non-zero $v$, we see that $\delta \QK \approx  - \frac{I_0}{2a^3} v^2$ and
\begin{align}
\delta_K \approx&  - \frac{1}{2} v^2 + \Order{I_0/a^3}. 
\end{align}

Considering now the dipole and setting $T^{(\rm K)}_{03} + T^{(f)}_{03}=0 $ we find 
\begin{equation}
\beta  = \frac{\rhobK}{a (1+w)\rhob}    \varphi_1, 
\end{equation}
to the lowest order approximation. Using our solution to write {$\beta$ in terms of $\delta_K$ gives
$\beta  =  \frac{\rhobK}{(1+w)\rhob}  \sqrt{2 |\delta_K|}$, 
so that in the late Universe
\begin{equation}
\beta  = \frac{\Omega_{0,K}}{(1+w)\Omega_{0,f} }   \sqrt{2|\delta_K|}. 
\label{khronondipole}
\end{equation}

Finally, the evolution of shear is given by
\begin{align}
  \dot{\sigma}  + 3 H \sigma    &= \kappa  \left[ \rhob (1 + w) \beta^2  +   \rhobK v^2 \right], 
\end{align}
and thus, in the late Universe, for a matter component $w=0$,
\begin{align}
  \dot{\sigma}  + 3 H \sigma    =&    \frac{3 H_0^2 \Omega_{0,K}\left(\Omega_{0,K} + \Omega_{0,f} \right)}{\Omega_{0,f} } \frac{\varphi_1^2 }{a^5}, 
\end{align}
which has the solution
\begin{align}
   \sigma =  \frac{1}{a^3} \left[  C +   \frac{3 H_0^2 \Omega_{0,K}\left(\Omega_{0,K} + \Omega_{0,f} \right) }{\Omega_{0,f} }   \varphi_1^2 \int^t \frac{dt'}{a^2} \right]. 
\end{align}
where $\Omega_{0,K}$ and $\Omega_{0,f}$ are the relative densities of the Khronon and fluid today, respectively.
The additional term now decays more quickly than the $a^{-3}$ of the conventional, homogeneous solution.

\section{Ancillary constraints}
\label{constraints}

All of the expressions found so far in the case of a small dipole show that $\beta$ is directly related to other, potentially observable, quantities. The exercise is now to look at how these other quantities are constrained by observations and how that impacts constraints on $\beta$.

\subsection{Curvature}

In an a curved anisotropic universe, we found that the dipole is related to the curvature and shear of the Universe through:
\begin{align}
    \beta \sim \frac{1}{1+w}{\Omega^{1/2}_{k}} \frac{\sigma_0}{H_0} 
\end{align}
where we assume $\Omega_0=\Omega_{0,f}\simeq 1$ and $H_0=100 h\, {\rm km} \, {\rm s}^{-1}\, {\rm Mpc}^{-1}$. 
Over the past two decades, constraints on the geometry of the Universe have improved dramatically. The main constraint comes from the positions of the peaks in the angular power spectrum of the cosmic microwave background (CMB) -- they serve as a a measurement of the angular diameter distance at last scattering -- combined with late time distance measurements from, for example, type-IA supernovae or baryon acoustic oscillations (BAOs) \cite{dodelson2020modern}. 

Current constraints from the CMB, using the 2018 data release of temperature and polarization data taken by the \textit{Planck} space observatory \cite{Planck:2018vyg}, 
$\Omega_k=-0.044^{+0.018}_{-0.015}$, indicating a very mild preference for curvature. Once CMB weak lensing is added, one finds 
$\Omega_k=-0.0106\pm 0.0065$ and including the (then) best measurements of the BAOs leads to
$
\Omega_k=-0.0007\pm 0.0019 $.
Thus we have a conservative upper bound on curvature, 
$
|\Omega_k|\lesssim  {\rm few} \times 10^{-3} 
$.

These constraints are obtained under the assumption that the Universe is homogeneous and isotropic and one might, therefore, be concerned about self consistency and applying them in this situation. But, given that we are interested in a scenario where deviations from isotropy are small -- recall that $\beta\simeq {\rm few} \times 10^{-3}$ -- we expect the effects on the parameter constraints to also be small compared to the overall effect of the background evolution. In other words, we expect percent level effects on the cosmological parameters. 

Global shear will also affect the CMB. As originally shown in \cite{Barrow:1985tda}, different Bianchi models will lead to additional patterns on the CMB which will be super-posed on the imprint of primordial perturbations: Bianchi I will lead to an additional quadropole (but no additional dipole) while Bianchi V will lead to a hot spot (with its size set by the curvature scale) and an additional dipole. In \cite{Bunn:1996ut} it was shown that this could be used to place constraints on shear today, of order
$
{\sigma_0/}{H_0}< {\rm few} \times 10^{-9} 
$.
Subsequently, this work was extended to include polarization \cite{Saadeh:2016sak} and the constraints from the Planck 2013 data on vorticity, $\omega/H_0$ translate to a vastly improved constraint on shear
$
{\sigma_0}/{H_0}< {\rm few} \times 10^{-12} 
$. 

Combining the constraints on curvature and shear we have that, at best, we can have
\begin{align}
\beta \sim  \frac{1}{1+w} \times 10^{-13}.
\end{align}
As we can see, we do not have much leeway to increase the amplitude of the dipole. Recall that $w$ is the equation of state of the boosted matter component. If we accept the observations, this means that we are talking about galaxies which are tied to the dark matter. Thus $w=0$ and $\beta\sim 10^{-13}$; approximately 10 orders of magnitude different from the observed dipole.

An alternative is to be more lax about what the boosted matter component {\it is} and relax our assumptions about its equation of state \cite{Krishnan:2022qbv}. By picking an equation of state that, very rapidly, approaches $w\simeq -1$ it is possible to enhance the dipole. One proposal has been to assume that $1+w\sim e^{-\lambda t}$ which approaches $0$ for large $t$ \cite{Krishnan:2022qbv}. This will have, again, knock on effects on other observables: the luminosity and angular diameter distances as a function of redshift constraints on the expansion rate of the Universe can, in turn, be used to place constraints on how rapidly the equation of state of the matter can deviate from that of non-relativistic matter \cite{DES:2024jxu,DESI:2025zgx} in the presence of an additional, dark energy component. An interesting aspect of this proposal is that, the more quickly $w$ transitions to $-1$, the more effective it will be at generating the dipole and the {\it less} constrained it will be by distance measurements (as it will only deviate from the conventional $\Lambda$CDM evolution very rapidly at very late times). Whether it is possible to construct a {\it bona-fide} physical model for such an equation of state is an open question.

\subsection{Heat flow}
In the setup we are considering, we took one species `A' to have velocity miscalligned from the usual FLRW cosmic flow and assumed that this is balanced by 
heat flow. This is a statement independent of which frame we may choose. We may indeed boost to a new frame (keeping always small $\beta$) 
such that the energy-momentum tensor of `A' becomes standard without any $q_\mu$ term.
However, in that frame a dipole will be generated in the other species and, moreover, the heat flow vector will remain invariant; see~\cite{Israel:1979wp,Kopp:2016mhm}.

If, indeed, there is a heat flow, it will be generated by a temperature difference along the dipole axis. 
From \eqref{Fourier_law} and assuming  $\Omega_{f0} \sim 0.3$, we find that, for a temperature difference $\Delta T$ across a distance $\Delta L$,
\begin{align}
\frac{\beta}{ 2.5 \times 10^{-3} } \sim&  
 \frac{2}{(1+w)h^2} \times 10^{-18}  
 \left(\frac{ \kappath}{ \qty{115}{\watt \cm^{-1} \kelvin^{-1}}  }\right) 
\nonumber
\\
&
\times 
 \left(\frac{\Delta T}{\qty{3}{\kelvin}}\right) \left(\frac{3000~\Mpc}{\Delta L}\right),
\label{Tcond}
\end{align}
where $\kappath = \qty{115}{\watt \cm^{-1} \kelvin^{-1}}$ is the thermal conductivity of  silver at a temperature of $\qty{3}{\kelvin}$.\footnote{
For reference, the thermal conductivity of silver at $\qty{3}{\kelvin}$ is $\qty{115}{\watt \cm^{-1} \kelvin^{-1}}$
and after a peak around $\sim\qty{7}{\kelvin}$ it falls to $\sim \qty{4.29}{\watt \cm^{-1} \kelvin^{-1}}$ around room temperature~\cite{Thermal_Conductivities}.
}  

We can use \eqref{Tcond} to speculate on the consequences of a large scale heat flow which would be required to generate a dipole. To begin with, if we assume that the flow is in the standard model sector, then the temperature gradient required will be imprinted on one, or all its ingredients. We have remarkably strict constraints on large scale temperature gradients in the CMB, either in the form of a constraint on its dipole or the strict upper bounds on spectral distortions over a range of frequencies \cite{1996ApJ...473..576F}. This leaves us with a few options. If the heat flow is being sourced by a CMB dipole which is subdominant to the kinematic dipole due to our motion relative the Sun, then one can only have a large enough heat flow if $\kappath$ is $\sim 10^{18}$ larger than that of silver.
The challenge is then to find a microphysical explanation for such a high thermal conductivity and what the subsidiary consequences might be. Another possibility is that the gradient is in the temperature of one of the other, standard model, cosmological fluids. Such fluids would have been coupled to the photon fluid (and radiation, more generally) but might have decoupled at very early times. The challenge then is to explain how such a gradient would have persisted in the decoupled fluid but would have been suppressed or erased in the CMB. This is not impossible -- causality limits the ability to generate large scale correlations but not to destroy them -- but a viable mechanism remains to be proposed.

The alternative is to consider the dark sector. It has been argued that the dark sector is far richer than had previously been considered with the possible existence of multiple, interacting species mirroring, in some sense, the standard model \cite{Ackerman_2009}. In that scenario, there is far more freedom in the microphysics and we cannot apply the usual lens to select ``viable'' scenarios. Nevertheless, the question remains of how one might imprint a large scale temperature gradient, or large thermal conductivity. Overall, regardless of the mechanism,
the thermal conductivities required are several orders of magnitude larger than  thermal conductivities of known materials or states of matter.

Finally, we note that the evolution of shear differs from that in an untilted Bianchi model; there is a correction which depends on the time evolution of $\beta(a)$. 
Depending on this time evolution there might be tighter constraints coming from, for example, the CMB.
 Lacking a complete microphysical discussion of $\beta(a)$ and a more specific underlying heat flow model, we refrain from using these constraints.

\subsection{Electromagnetic Field}

We have that, in a universe with large-scale electric and magnetic fields, we can express the resulting dipole in a convenient form 
\begin{align}
\beta 
 \sim& \frac{15}{h^2}\frac{1}{1+w} \left(\frac{B_0}{10^{-3} \ {\rm Gauss}}\right) \left(\frac{E_0}{ 10^{-3}\  {\rm Statvolt/cm}}\right)
\end{align}
where the fields are in CGS units. 

The first thing to note is the dependence on the electric field, which is usually absent in any cosmological setting. First, if we assume that the electrical field is the one arising in the electromagnetic sector of the standard model, one then has to take into account the value of the conductivity, $\sigma_c$, of the Universe. At very early times (before electron-positron annihilation) one has that $\sigma_c\sim T$, i.e. it decays with time \cite{Caprini:2003gz}. During the radiation era, $\sigma_c=(2\pi m_e/e^2)(n_e/n_\gamma)\sim 10^{13} {\rm s}^{-1}$, where $m_e$ is the mass of the electron and $n_e/n_\gamma$ is the ratio of the electron to photon density. In the matter era, one is left with a residual distribution of photons and, through Thomson scattering with CMB photons, one is left with $\sigma_c \sim 10^{10} {\rm s}^{-1}$. Thus, $\sigma_c$ is very large and, in the ideal magnetohydrodynamic limit, assuming finite currents and Ohms law, ${\vec J}=\sigma_c {\vec E}$, the electric field will go to zero. Thus, the usual electromagnetic field cannot support a dipole.

As mentioned above, it has been conjectured that the dark sector may be richer than had previously been considered. One aspect of this is that a whole ``dark gauge'' Universe might exist with associated gauge fields and charged dark matter that would interact under these gauge fields. In that case, if one relaxes any assumptions about the conductivity of the dark sector, it is possible to support large, dark electrical fields. Nevertheless, it is still possible to obtain constraints on large-scale, dark magnetic fields through their impact on the anisotropic expansion of the Universe. As was shown in \cite{Barrow:1997mj}, in the case of no electric field, one has that $\sigma/{H} \sim {\rho_B}/{\rho}$ 
where $\rho_B$ is the energy density in the magnetic field and  
$\rho$ is the energy density in the remaining, dominant components. Note that, for this particular case there is an additional, steeply decaying mode for the shear which may enhance the anisotropy at early times.  Thus, using current bounds on $(\sigma/H_0)$ we have a conservative bound of $
B_0<\Omega_010^{-10} h^2
$.
Saturating this bound, and recalling that $E_0=B_0$, we have that 
\begin{align}
\beta \sim &   \frac{\Omega_0}{1+w}\ \times 10^{-8}.
\end{align}
Thus, it would seem challenging to generate a dipole of the desired amplitude using large-scale electromagnetic fields.

\subsection{Khronon}
In a universe that supports a Khronon field, we found that the dipole today is related to the Khronon density perturbations via
\begin{equation}
\beta=\frac{\Omega_K}{(1+w)\Omega_f}\sqrt{2|\delta_K|}, 
\end{equation}
where, again, we have relaxed the assumption that matter is non-relativistic, allowing $w\neq 0$. We also assume that $\Omega_f$
is composed of baryons and part of dark matter and so it can not vanish fully, however, it could be as small as the baryon relative density if no dark matter follows 
the dipole.
 As in the previous cases, $w\rightarrow -1$ will boost the dipole but it will clash with the abundance of evidence for non-relativistic matter in a number of cosmological observables, most notably, in measurements of the expansion rate \cite{DESI:2025zgx}.

It is difficult to estimate the magnitude of $\delta_K$ and it will depend on the nature and source of perturbations. Yet, if one assumes 
\emph{adiabatic} initial conditions, one has $\delta_K\sim \delta_{M}$ and thus, on horizon scales, we expect $\delta_K\simeq 10^{-5}$. 
If $\Omega_K\sim \Omega_f$ then the amplitude of the dipole in a Khronon universe is tantalizingly close to the reported observed value. If $\Omega_f$ is just the baryon fraction of matter then $\Omega_K/\Omega_f$ can be as large as a factor of five if the Khronon provides the totality of dark matter, which would seem to make
this possibility stronger. However, in this extreme case,  during matter domination, one can boost into a frame where there is no dipole, hence, the Khronon can only 
constitute part of dark matter if this is to work. A natural choice would be equipartition, such that the baryon plus Khronon densities
constitute roughly half of the matter density, so that $\Omega_K/\Omega_f \sim 2$. The issue then shifts to explaining why this ratio takes such values.
Having much smaller values leads to  $\beta\ll 10^{-3}$. 

Alternately, one can assume  isocurvature fluctuations, with $\rho_f\delta_f+\rho_K\delta_K\sim 0$ on large scales, and assume that 
$\Omega_K\ll\Omega_f$. We are then in a situation in which we can have $\delta_K\gg 10^{-5}$ which might compensate for the small $\Omega_K$. We then have
\begin{eqnarray}
    \beta_{\rm iso}\simeq 
\sqrt{2\frac{\Omega_K}{\Omega_f}|\delta_f|}\sim 10^{-2}\times \sqrt{\frac{\Omega_K}{\Omega_f}} 
\end{eqnarray}
If $\Omega_K$ is subdominant, then the dipole will be suppressed. Having a larger $\Omega_K$ would then lead to substantial matter isocurvature contribution to the 
CMB angular power spectrum, on which we have strong constraints~\cite{Bucher:2004an,Dunkley:2005va,Planck:2018jri}.

Constraints on the shear can play a role here. For simplicity, if we asssume matter domination, we have that the correction to the shear will be of the form $\sigma \simeq 2\delta_K H_0 a^{-7/2}$, plus the homogeneous solution $\sim a^{-3}$. 
This means that we can place, from constraints on $\sigma_0/H_0$, tight constraints on $\delta_K$ where $\delta_K\ll 10^{-5}$. This would, again, void the possibility of having a large enough anomalous dipole. We leave a more detailed exploration of the role of shear in Khronon cosmologies for later work.

\section{Discussion}
\label{discussion}
In this paper we have constructed models of the Universe with relative bulk motion of matter in the CMB frame, as suggested by the quasar dipole anomaly {under a purely kinematic interpretation}.
We have picked a particular class of models -- tilted Bianchi models -- which are, to some extent, manageable and in which we can find explicit expressions for the {kinematic} dipole as a function of other physical properties. This has allowed us illustrate what we believe to be a key aspect of any attempt to construct a theoretical understanding of an anomalous cosmological dipole: that there should be ancillary observational consequences which may be probed by a variety of different techniques. This makes such models testable and we find that they all fail. 

We have picked four scenarios which, intuitively, can lead to a dipole: curvature, heat flow, electromagnetic field and the khronon field. Of these scenarios, three will all lead to egregious anomalous observational effects in other probes. Thus it is possible to place bounds on the dipole via other means which make the current observed value prohibitively large. In the case of the Khronon field, one is tantalizingly close: a more detailed modelling of the linearly perturbed Khronon Universe is required before we can make definitive statements about the viability of such a large dipole.

One aspect that we have (partially) explored is that the presence of the dipole, at least in homogeneous anisotropic models, leads to a modification in global shear which itself can be constrained by other means. An interesting question is whether this symbiotic relationship between dipole and shear will be present more generally in other models with even less symmetry than the ones we considered here. A way of addressing this would be the covariant formalism of \cite{ellis2012relativistic} to extract general constitutive relations between the dipole, shear and any other relevant degrees of freedom. 

An important aspect that we have not explored here is the role of the very early Universe in these models. It is well known that inflation is endowed with a ``no-hair'' theorem that precludes the existence of large global anisotropies at early times \cite{Wald:1983ky}. Given that these anisotropies generally decay in time, it means that it would be practically impossible to have large effects of the size we need to support such a large observed dipole. Thus, if the observed dipole is real, and we can describe the large scale structure of the Universe as (roughly) a globally anisotropic model, then have we uncovered a profound new insight on the early Universe and the death knell for inflation? This remains to be seen, given existing attempts at incorporating anisotropic expansion into inflationary model building.

Does the fact that almost all the models we have considered fail other observational tests mean that attempts at constructing a theoretical model for the cosmological dipole are doomed? Not necessarily. We have picked a restricted class of models, the simplest extensions to FLRW that one might consider but in which one can incorporate a dipole. The fact that they are so simple is why they are so constraining. The lack of freedom to accommodate any type of observation is a desirable property in a predictive model but it may happen that slightly more complex models will be observationally viable. For example, it may be possible to tailor simple inhomogeneous models, such as the Lemaitre-Tolman-Bondi model \cite{Lemaitre:1933gd,Tolman:1934za,Bondi:1947fta}, to accommodate a dipole.
But there is a concern that, once one starts constructing ever more elaborate models, one begins to lose much of the predictive power of contemporary cosmology. And if that is the case, it becomes unclear where one should stop, or how successful the cosmological programme has been. The role of an anomalous dipole in the future of cosmology remains to be seen.

\section*{Acknowledgements}

We thank George Ellis, Sebastian von Hausegger and Roy Maartens for useful discussions and comments on the draft.

AM acknowledges support financial support for a DPhil studentship from the Department of Physics at Oxford. PGF acknowledges support from STFC and the Beecroft Trust. 
DJB acknowledges that support was provided by Schmidt Sciences, LLC.
HD is supported by a Royal Society University Research Fellowship (grant no. 211046).
CS acknowledges support from the European Structural and Investment Funds and the Czech Ministry of Education, Youth and Sports (project No. FORTE---CZ.02.01.01/00/22\_008/0004632)  and from the Royal Society Wolfson Visiting Fellowship ``Testing the properties of dark matter with new statistical tools and cosmological data''. TY acknowledges the support of a UKRI Frontiers Research Grant [EP/X026639/1], which was selected by the European Research Council.

For the purpose of open access, the authors have applied a Creative Commons Attribution (CC BY) licence to any Author Accepted Manuscript version arising.

\bibliography{refs}

\end{document}